\documentclass[11pt]{amsart}
\usepackage{amscd}
\usepackage{amsmath}
\usepackage{graphicx}
\usepackage{amsfonts}
\usepackage{amssymb}
\usepackage{epstopdf}
\begin{document}
\title[Approximate separation of quantum gates]
{Approximate separation of quantum gates and separation experiments
of CNOT based on Particle Swarm Optimization algorithm}
\author{Kan He$^1$, Shusen Liu$^2$, Jinchuan Hou$^1$}
\address {1. College of
Mathematics, Taiyuan University of Technology, Taiyuan 030024, P. R.
China; 2. Institute for Quantum Computing, Baidu Research, Beijing
100193, P. R. China} \email[K. He]{hekanquantum@163.com}


\begin{abstract}

 Ying conceived of using two or more small-capacity quantum computers to produce
    a larger-capacity quantum computing system by quantum parallel programming ([M. S. Ying,  Morgan-Kaufmann, 2016]).
    In doing so, the main obstacle
    is separating the quantum gates in the whole circuit to produce a tensor product of the local gates. It has been showed  that there are few separable
    multipartite quantum gates, so the approximate separation problem involves finding local quantum gates that
    approximate a given inseparable gate. We propose and study a problem involving the approximate separation
    of multipartite gates based on quantum-gate fidelity. For given multipartite and local gates, we conclude that
    the smaller is the maximal distance between the products of an arbitrary pair of eigenvalues, the greater is
    their gate fidelity. This provides a criterion for approximate separation. Lastly, we discuss the optimal
    approximate separation of the CNOT gate.

\end{abstract}
\maketitle

\section{Introduction}

Programming for quantum computers has become an urgent task nowaday
\cite{sof}, \cite{qlanguage}, \cite{mue}, \cite{zeng}. As reported
in \cite{sof}, \cite{sel}, \cite{gay}, \cite{yingbook}, extensive
research has been conducted on quantum programming over the last
decade, and several quantum programming platforms have been
developed over the last two decades. The first quantum programming
environment was the `QCL' project proposed by \"{O}mer in
1998~\cite{qcl1}, \cite{qcl2}. Then, more quantum softwares are
emerged, for instance,  Q language as a C++
library~\cite{qlanguage}, a scalable functional quantum programming
 language, called Quipper ~\cite{green}, ~\cite{Scafford}, ~\cite{scaffCC}. Wecker and Svore from QuArc (the Microsoft Research Quantum Architecture and Computation team) developed
   LIQU$i|\rangle$ as a modern tool-set embedded within F\#~\cite{Liquid}.
At the end of 2017, QuARC announced a new programming language and
simulator designed specifically for full-stack quantum computing,
known as Q\#, which represents a milestone in quantum programming.
 In the same year, Liu et al. released the quantum program Q$|SI\rangle$  that supports a more complicated loop structure~\cite{liu2017q}.
 To date, the structures of programming languages and tools have mainly been sequential.

Beyond the constraints of quantum hardware,
 there remain several barriers to the development of practical
  applications for quantum computers. One of the most serious
   barriers is the number of physical qubits provided in physical machines.
    For example, IBMQ produces two five-qubits quantum computers ~\cite{ibmqx2} and one
    16-qubit quantum computer ~\cite{ibmqx3}, which are available to programmers through the cloud,
     but these are far fewer qubits than are required by practical quantum algorithms.
Today, quantum hardware is in its infancy. As the number of
available qubits is gradually increasing, many researchers are
considering the possibility of combining various quantum hardware
components to work as a single entity and thereby enable advances in
the number of qubits~\cite{yingbook}. To increase the number of
accessible qubits in quantum hardware, one approach uses concurrent
or parallel quantum programming. Although current quantum-specific
environments are sequential in structure, some researchers are
working to exploit the possibility of parallel or concurrent quantum
programming on the general programming platform from different
respects. Vizzotto and Costa applied mutually exclusive access to
global variables to enable concurrent programming in
Haskell~\cite{vizzotto2005concurrent}. Yu and Ying
 studied the termination of concurrent
programs~\cite{yu2012reachability}. Researchers provide mathematics
tools for process algebras to describe their interaction,
communication and synchronization~\cite{gay2005communicating}.
Recently, Ying and Li  defined and established operational
(denotational) semantics and a series of proof rules for ensuring
the correctness of parallel quantum programs\cite{Yingli}. Not
surprisingly, we showed that multipartite quantum gates that can be
separated simply seldom exist \cite{HeLiuHou}. Furthermore, in a
practical quantum circuit, we must know how a given multipartite
gate can be closed by local gates. In this paper, we show that for
given multipartite and local gates, the smaller is the maximal
distance between the products of an arbitrary pair of eigenvalues,
the greater is their gate fidelity. This provides a criterion of
approximate separation. We also discuss the optimal approximate
separation of the CNOT gate.

It has been showed that only few kinds of multipartite gates can be
separated directly. In this section, we turn to study the
approximate separation problem of multipartite gates.

We introduce some notations. Let ${\mathcal H}_k$ be a complex
Hilbert space with dim$H=m_k$, and $ \otimes_{k=1}^n {\mathcal H}_k$
the tensor product of ${\mathcal H}_k$s.
 Still denote by $ \mathcal B(\otimes_{k=1}^n {\mathcal H}_k), \mathcal U(\otimes_{k=1}^n {\mathcal H}_k)$ and $\mathcal B_s(\otimes_{k=1}^n {\mathcal H}_k)$
 the set of all bounded linear operators, all unitary operators, and all self-adjoint operators
  on the underline space $\otimes_{k=1}^n {\mathcal H}_k$ respectively.
The error of two gates $U, V$ is
$$E(U,V)={\rm max}_{|x\rangle}\{d(U |x\rangle\langle x| U^\dagger,
V|x\rangle\langle x| V^\dagger): \||x\rangle\|=1\},$$ where $d$ is
an arbitrary distance between two matrices.

\section{Results and experiments}

{\bf The problem on $\epsilon$-approximate separation} {\it Given a
positive scalar $\epsilon$ and a multipartite quantum gate
    $U\in \mathcal U(\otimes_{k=1}^n {\mathcal H}_k)$, determine whether there are local gates
    $U_i\in {\mathcal U}({\mathcal H}_i)$ such that
    \begin{equation}
    \label{eq:2.1}
    E(U, \otimes_{i=1}^n U_i)\leq \epsilon,
    \end{equation}
    where $d(\cdot,\cdot)$ is an arbitrary distance between two operators. We refer to $U$ as the $\epsilon$-approximate separable one if Eq.~\ref{eq:2.1} holds true.
    Furthermore, how do we find these local gates $U_i$?} \\

In our solution, the gate fidelity is selected as the replacement of
the distance between matrices in the above question. The gate
fidelity between two unitary gates $U, V$ is defined as
$$F_{\rm min}(U, V)={\rm min}_{|x\rangle}\{F(U|x\rangle \langle x|
U^\dagger, V|x\rangle\langle x| V^\dagger): \||x\rangle\|=1\},$$
where $F(A,B)={\rm tr}(\sqrt{\sqrt{A}B\sqrt{A}})$ is the Uhlmann
fidelity. Considering the specialty of the gate fidelity, we
re-describe the question on $\epsilon$-approximate separation as
follows:

{\bf The question on $\epsilon$-approximate separation  based on the
gate fidelity} {\it Given a positive scalar $\epsilon$ and a
multipartite quantum gate
    $U\in \mathcal U(\otimes_{k=1}^n {\mathcal H}_k)$, whether or not there are local gates
    $U_i\in {\mathcal U}({\mathcal H}_i)$ such that
    \begin{equation}
    \label{eq:3.2}
    F_{\rm min}(U, \otimes_{i=1}^n
    U_i)\geq 1-\epsilon.
    \end{equation}
    And how do we find these local gates $U_i$?} \\

Next we answer the question on $\epsilon$-approximate separation
based on the gate fidelity by connecting the gate fidelity to
numerical ranges. Recall the numerical range of a bounded linear
operator $A$ is
$$W(A)=\{\lambda=\langle x|A|x\rangle, \||x\rangle\|=1\}.$$ The
numerical radius of $A$ is
$$w(A)={\rm max} \{|\lambda|: \lambda\in W(A)\}.$$ Let us introduce
the distance from 0 to $W(A)$ is defined as $$w_{\rm min}(A)={\rm
min}\{|\lambda|:\lambda\in W(A)\}.$$   The maximal distance of the
eigenvalues of a matrix $A$ with its spectral set $\sigma(A)$:
$$d_{\rm max}(A)={\rm max}_{\lambda_i\in \sigma(A)}\{|\lambda_i-\lambda_j|\}.$$

{\bf Theorem 2.1} {\it For arbitrary $m\times m$ unitary matrices
$U,V$,
$$F_{\rm min}(U,V)=\sqrt{1-(\frac{d_{\rm
max}(V^\dagger U)}{2})^2}.$$}

{\bf Remark. } From Theorem Theorem 2.1, we conclude that the
$\epsilon-$-approximate separation question that is based on the
gate fidelity can be solved in the following manner: for a given
multipartite quantum gate $U\in \mathcal U(\otimes_{k=1}^n {\mathcal
H}_k)$, we design a search program to find the unitary matrix
    $U_i\in {\mathcal U}({\mathcal H}_i)$ for each $i$ such that $d_{\rm max}((\otimes_{i=1}^n
U_i^\dagger) U)$ converges to its infimum. Simultaneously, the gate
fidelity $F_{\rm min}(U, \otimes_{i=1}^n
    U_i)$ can touch its  its supremum. Therefore, $ \otimes_{i=1}^n U_i$ can become an $\epsilon-$approximate
separation to $U$ based on the gate fidelity if and only if $d_{\rm
max}(\otimes_{i=1}^n U_i^\dagger U)\leq
2\sqrt{2\epsilon-\epsilon^2}$ (assume without loss of generality
$\epsilon< 1$).

{\bf Proof of Theorem 2.1. } First, it is easy to check that
$$F_{\rm min}(U, V)={\rm min}_{\||x\rangle\|=1}|\langle x|V^\dagger
        U|x\rangle|=w_{\rm min}(V^\dagger U).$$
Note that the numerical range of each an $m\times m$ unitary matrix
is a convex polygon with its $m$ vertex lying on the circumference
of the disc with the unit radius. It follows that the value $w_{\rm
min}(V^\dagger U)$ equals to the distance from the original point to
the longest edge of the convex polygon. We therefore have that
$$w_{\rm min}(V^\dagger U)=\sqrt{1-(\frac{d_{\rm max}(V^\dagger
U)}{2})^2}.$$  Thus, we complete the proof.  \hfill$\square$

We can understand this through Figure 1, where we describes
the $m=4$  case. Note that where all eigenvalues of $V^\dagger U$
lie on the circumference of the circle with the unit radius.

\begin{figure}[!htb]
\includegraphics[width=2.9in]{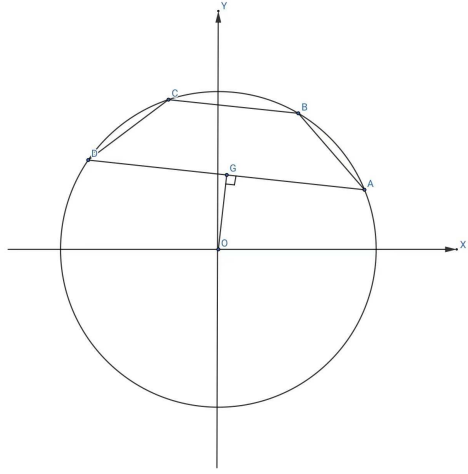}
\caption{ ${\rm A, B, C, D}$ are eigenvalues of $V^\dagger U$. The
quadrangle ${\rm ABCD}$  is the numerical range $W(V^\dagger U)$.
The line segment ${\rm AD}$ is $d_{\rm max}(V^\dagger U)$.
Furthermore, the line segment $OG$ equals to $w_{\rm min}(V^\dagger
U)$, which is $F_{\rm min}(U, V)$. It is clear from the graph that
the theorem holds true. }
\end{figure}

{\bf Example 2.2} Applying the above theoretical analysis, we
consider the approximate separation question of CNOT gates. The CNOT
gate is of the following form: ${\rm \bf CNOT}=\begin{pmatrix} 1 & 0
& 0 & 0 \cr 0 & 1& 0 & 0 \cr 0 & 0 & 0 & 1 \cr 0 & 0 & 1 & 0
\end{pmatrix}$. Assume that $U=\begin{pmatrix} u_{11} & u_{12} \cr u_{21} & u_{22}
\end{pmatrix}$ and $V=\begin{pmatrix} v_{11} & v_{12} \cr v_{21} & v_{22}
\end{pmatrix}$ are two $2\times 2$ unitary matrices. Since a $2\times 2$ unitary matrix with its four real parameters can be represented in the form
 $$e^{i\alpha }\begin{pmatrix} e^{-i\beta} & 0 \cr 0
& e^{i\beta}
\end{pmatrix}\begin{pmatrix} \cos \frac{\gamma}{2} & -\sin \frac{\gamma}{2} \cr \sin \frac{\gamma}{2} &
\cos \frac{\gamma}{2}
\end{pmatrix}\begin{pmatrix} e^{-i\delta} & 0 \cr 0 &
e^{i\delta}
\end{pmatrix}, $$ we denote $U, V$ by $U[\alpha_U, \beta_U, \gamma_U,
\delta_U]$ and $V[\alpha_V, \beta_V, \gamma_V, \delta_V]$
respectively.

We find the $\epsilon$-approximate separation solution of ${\rm\bf
CNOT}$ by the following optimization problem:

{\bf Maximize}: $d_{\rm Max}(U^\dagger\otimes V^\dagger{\rm\bf
CNOT})$

{\bf subject to}: $U, V$ are 2$\times$2 unitary matrices\\
Since  $U=U[\alpha_U, \beta_U, \gamma_U, \delta_U]$ and
$V=V[\alpha_V, \beta_V, \gamma_V, \delta_V]$, $d_{\rm
Max}(U^\dagger\otimes V^\dagger{\rm\bf CNOT})$ is a nonnegative
function with eight parameters $f_m[\alpha_U, \beta_U, \gamma_U,
\delta_U, \alpha_V, \beta_V, \gamma_V, \delta_V]$. Thus, the above
optimization problem is equivalent to:

{\bf Maximize}: $f_m[\alpha_U, \beta_U, \gamma_U, \delta_U,
\alpha_V, \beta_V, \gamma_V, \delta_V]$

{\bf subject to}: $\alpha_U, \beta_U, \gamma_U, \delta_U, \alpha_V,
\beta_V, \gamma_V, \delta_V\in {\Bbb R}$\\

Applying the particle swarm optimization algorithm, we obtain the
optimal output as $d_{\rm Max}(U^\dagger\otimes V^\dagger{\rm\bf
CNOT})\approx 1.4159$, $U\approx U[218.0000, 157.0000, 159.0000,
471.0000],$ \\ $ V\approx V[633.0000, 84.0000, 628.0000, 387.0000].$
From Theorem 2.1, it follows that the corresponding value of the
gate fidelity is approximately $0.7063$. At the moment, the two
unitary matrices that approximately separate the CNOT gate with
$\epsilon\approx 1-0.7063=0.2937$ are in form
$$U=\begin{pmatrix} 0.4057-0.5795i & 0.5800+0.4040i \cr
0.5793+0.4049i & 0.4039-0.5808i
\end{pmatrix}$$ and $$V=\begin{pmatrix} 0.6724+0.7402i &
-0.0016i \cr 0.0002-0.0016i & 0.7386-0.6741i
\end{pmatrix}.$$


Next we make a numerical experiment on the above approximate
separation shown above.  Each one among 1000 random pure two-qubit
states $|\psi\rangle $ is acted upon by ${\rm\bf CNOT}$ and the
tensor product $U\otimes V$ respectively. In Figure 2, for an
arbitrary random two-qubit input  $|\psi\rangle $, the values of the
y-axis represent the Unhlmann fidelity of ${\rm\bf
CNOT}(|\psi\rangle)$ and $U\otimes V(|\psi\rangle)$. It can be seen
that the values of the y-axis is bounded from below by 0.7063, and
part of them values are close to 1.

\begin{figure}[!htb]
\includegraphics[width=2.9in]{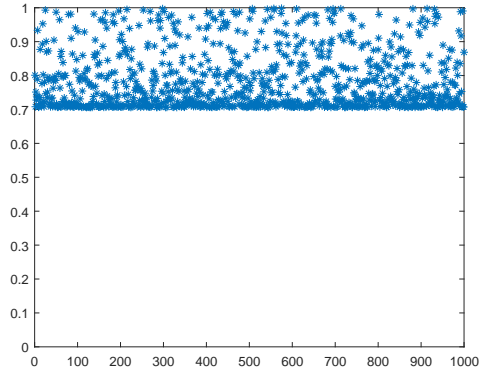}
\caption{The numerical experiment on the approximate separation of
{\rm\bf CNOT}. }
\end{figure}

\section*{Conclusion and discussion}

In this paper, we proposed and discussed the
 approximate separation question, which has more practical import.
Here we identified an interesting connection between gate fidelity
and the numerical range of operators, and solved the approximate
separation question of multipartite quantum gates based on the gate
fidelity. Furthermore, we provided an example of the approximate
separation of the CNOT gate.

{\bf Acknowledgements} Thanks for comments. This work is partly supported by National
Natural Science Foundation of China No. 11771011, 12071336.
Correspondence  should be addressed to  K. He(hekanquantum@163.com).

\end{document}